\documentclass[12pt,american]{article}
\usepackage[T1]{fontenc}
\usepackage[latin9]{inputenc}
\usepackage{fancyhdr}
\usepackage{verbatim}

\makeatletter

\date{\mbox{\small February 16, 2013}}

\usepackage{theorem}

\theorembodyfont{\upshape }
\theoremheaderfont{\itshape }

\newtheorem{theorem}{\null\hspace*{1.\parindent}Theorem}

\newtheorem{definition}{\null\hspace*{1.\parindent}Definition}

\def\section{\setcounter{equation}{0}\@startsection {section}{1}{\z@}
{-3.5ex plus -1ex minus -.2ex}{2.3ex plus .2ex}{\normalsize\bf}}

\newcommand{\tochka}{\hspace*{-0.9ex}. \ }

\makeatother

\usepackage{babel}

\begin{document}

\author{Larissa A. Manita~%
\thanks{This research was supported by the Russian Foundation for Basic Research,\protect\\ 
project No.~11-01-00986-a~.%
} }

\title{~\protect\\*[-30mm] Optimal chattering solutions for longitudinal
vibrations of a nonhomogeneous bar\\
 with clamped ends }

\maketitle
\begin{center}
~\vspace{-1.6cm}

\par\end{center}

\begin{center}
\emph{National Research University Higher School of Economics,}\\
\emph{Moscow Institute of Electronics and Mathematics,}\\
\emph{Bolshoy Trehsviatitelsiy Per.~3,}\\
\emph{109028 Moscow, Russia}
\par\end{center}

\begin{center}
\texttt{\vspace{0.3cm}
lmanita@hse.ru}
\par\end{center}
\begin{abstract}
We consider a control problem for longitudinal vibrations of a nonhomogeneous
bar with clamped ends. The vibrations of the bar are controlled by
an external force which is distributed along the bar. For the minimization
problem of mean square deviation of the bar we prove that the optimal
control has an infinite number of switchings in a finite time interval,
i.e., the optimal control is the chattering control. 
\end{abstract}

\section{Introduction}

Consider small longitudinal vibrations of a nonhomogeneous bar of
length $l$. The longitudinal displacement at a typical point $x$
is denoted $y(t,x)$ where $t$ is the time. Let $g(x,t)$ be a density
of external longitudinal force at the instant of time $t$ at the
point $x$. Suppose that \[
g(t,x)=u(t)f(x)\]
where the force profile function $f(x)$ is assumed to be given, $u\left(t\right)$
is the control function. We assume that \begin{equation}
-1\leq u\left(t\right)\leq1\label{eq:bound_control-1}\end{equation}
The equation of longitudinal vibrations of the bar can be written
as \begin{equation}
p\left(x\right)y_{tt}(t,x)-\left(k\left(x\right)y_{x}(t,x)\right)_{x}=u(t)f(x)\label{eq:_1}\end{equation}
Here \[
p\left(x\right)=\rho\left(x\right)S\left(x\right),\quad k\left(x\right)=E\left(x\right)S\left(x\right)\]
where $\rho\left(x\right)$ is the density of bar, $S\left(x\right)$
is the cross-sectional area, $E\left(x\right)$ is the Young's modulus
at $x$, see, for example, \cite{Jov-Kosh-2012,Udwadia,Sadek}. 

We assume that the ends of the bar are clamped: \begin{equation}
\left.y\right|_{x=0}=\left.y\right|_{x=l}=0,\quad t>0\label{eq:gr_1}\end{equation}
 and the initial position and velocity are fixed: \begin{equation}
\left.y\right|_{t=0}=y_{0}(x),\quad x\in[0,l]\label{eq:na_us_1}\end{equation}

\begin{equation}
\left.y_{t}\right|_{t=0}=y_{1}(x),\quad x\in[0,l]\label{na_us_2}\end{equation}

We suppose that the coefficient functions $k$, $p$ are smooth enough
and \begin{equation}
\forall x\in\left[0,l\right]\,\,\, k\left(x\right)\geq k_{0}>0,\quad p\left(x\right)\geq p_{0}>0\label{usl:k_p}\end{equation}

We consider an optimal control problem: to find such a control function
$u\left(t\right)$ that minimize the following functional

\begin{equation}
\int_{0}^{\infty}\int_{0}^{l}p\left(x\right)y^{2}\left(t,x\right)dxdt\rightarrow\inf\label{functional_u^2}\end{equation}
under~(\ref{eq:_1})--(\ref{usl:k_p}). 

The problems of longitudinal vibrations of a bar were considered in
\cite{Jov-Kosh-2012,Udwadia,Sadek}. In~\cite{Jov-Kosh-2012,Udwadia}
the dynamics of the longitudinal vibrations of a bar subjected to
viscous boundary conditions was studied. The optimal boundary control
problem for longitudinal vibrations of a bar was considered in \cite{Sadek}.
By using a maximum principle the optimal control was expressed in
terms of an adjoint variable.

In this paper for the problem of controlling the longitudinal vibrations
of a bar~(\ref{eq:bound_control-1})--(\ref{functional_u^2}) we
construct a solution $y\left(t,x\right)$ in the form \begin{equation}
y(t,x)=\sum_{j=1}^{\infty}s_{j}(t)h_{j}(x)\label{eq:y}\end{equation}
 where $\left\{ h_{j}\left(x\right)\right\} _{j=1}^{\infty}$ are
eigenfunctions of the Sturm-Liouville problem, $\left\{ s_{j}\left(t\right)\right\} _{j=1}^{\infty}$
are corresponding Fourier coefficients. To find Fourier coefficients
we consider an optimal control problem in the space $l^{2}$. For
the control problem in $l^{2}$ we show that the optimal solutions
contain singular trajectories and chattering trajectories. A trajectory
is called a chattering trajectory if it has an infinite number of
a control switchings on a finite time interval. By similar method
we studied the optimal control problem for a rotating uniform Timoshenko
beam~\cite{zel_manita_2006,bor_zel_manita_2008}. But for the Timoshenko
beam similar results hold only for a dense set of initial conditions
in the space~$l^{2}$.

\section{Optimal control problem in $l_{2}$}

Define an operator $L$ in $C^{2}\left(\left[0,l\right]\right)$ by
\[
Lh=\left(kh{}_{x}\right)_{x}\]

Consider the following Sturm-Liouville eigenvalue problem with Dirichlet
boundary conditions: \begin{equation}
Lh+\lambda p\left(x\right)h=0,\quad x\in\left(0,l\right)\label{eq:SLP}\end{equation}
\begin{equation}
h\left(0\right)=0,\quad h\left(l\right)=0\label{gran_usl_SLP}\end{equation}

Here the functions $k\left(x\right)$ and $p\left(x\right)$ satisfy~(\ref{usl:k_p}).
It is known (see \cite{Gwaiz}) that the problem~(\ref{eq:SLP})--(\ref{gran_usl_SLP})
has an infinite sequence of eigenvalues $\left\{ \lambda_{j}\right\} _{j=1}^{\infty}$,
which are simple and positive: \[
0<\lambda_{1}<\lambda_{2}<\ldots,\quad\quad\lambda_{j}\rightarrow\infty,\quad j\rightarrow\infty\]
To each eigenvalue $\lambda_{j}$ corresponds a single eigenfunction
$h_{j}$, and the sequence of eigenfunctions $\left\{ h_{j}(x)\right\} _{j=1}^{\infty}$
forms an ortonormal basis of $L_{2}\left(\left(0,l\right);p\right)$
with the inner product $\left(z,w\right)_{p}=\int_{0}^{l}p\left(x\right)z\left(x\right)w\left(x\right)dx$.
If $k\left(x\right),p\left(x\right)$ are smooth enough (for example,
$k',\, p\in C^{1}\left([0,l]\right)$) and the condition~(\ref{usl:k_p})
holds, then the eigenvalues $\lambda_{j}$ admit the asymptotic form
\cite{zettl,Petrovsky,Collatz}: \begin{equation}
\frac{\lambda_{j}}{j^{2}}\sim\pi^{2}\left(\int_{0}^{l}\sqrt{p\left(x\right)/k\left(x\right)}dx\right)^{-2},\quad j\rightarrow\infty\label{asymp_lambda}\end{equation}

Assume that $y(t,\cdot)\in L_{2}\left(\left(0,l\right);p\right)$.
For any $t>0$ we expand the solution $y\left(t,x\right)$ of~(\ref{eq:_1})
in the basis $\left\{ h_{j}(x)\right\} _{j=1}^{\infty}$: \begin{eqnarray}
y(t,x) & = & \sum_{j=1}^{\infty}s_{j}(t)h_{j}(x)\quad\label{rjad_fur}\\
s_{j}(t) & = & \int_{0}^{l}p\left(x\right)y(t,x)h_{j}\left(x\right)dx\,=\,\left(y,h_{j}\right)_{p}\nonumber \end{eqnarray}
Using (\ref{eq:na_us_1})--(\ref{na_us_2}) we get: \[
s_{j}\left(0\right)=\int_{0}^{l}p\left(x\right)y_{0}\left(x\right)h_{j}\left(x\right)dx=\left(y_{0},h_{j}\right)_{p}\]
\[
\dot{s}_{j}\left(0\right)=\int_{0}^{l}p\left(x\right)y_{1}\left(x\right)h_{j}\left(x\right)dx=\left(y_{1},h_{j}\right)_{p}\]
 We multiply the equation (\ref{eq:_1}) by $h_{j}$ and integrate
it in $x$: \[
\int_{0}^{l}\left(py_{tt}+Ly\right)h_{j}dx=\int_{0}^{l}ufh_{j}dx\]
\[
\frac{d^{2}}{dt^{2}}\left(y,h_{j}\right)_{p}+\left(Ly,h_{j}\right)=u\left(f,h_{j}\right)\Rightarrow\quad\frac{d^{2}}{dt^{2}}\left(y,h_{j}\right)_{p}+\left(y,Lh_{j}\right)=u\left(f,h_{j}\right)\]
or \[
\frac{d^{2}}{dt^{2}}\left(y,h_{j}\right)_{p}+\lambda_{j}\left(y,h_{j}\right)_{p}=u\left(f,h_{j}\right)\]
Thus the function $s_{j}\left(t\right)$ satisfies the following equation:
\[
\ddot{s_{j}}(t)+\lambda_{j}s_{j}\left(t\right)=C_{j}u(t),\quad j=1,2,\ldots\]
where \begin{equation}
C_{j}=\left(f,h_{j}\right)=\int_{0}^{l}f\left(x\right)h_{j}\left(x\right)dx\label{eq:C-j}\end{equation}
 We substitute~(\ref{rjad_fur}) into~(\ref{functional_u^2}). Using
Parseval's equality  we get:\begin{equation}
\int_{0}^{\infty}\int_{0}^{l}p\left(x\right)y^{2}\left(t,x\right)dxdt=\int_{0}^{\infty}\sum_{j=1}^{\infty}s_{j}^{2}(t)dt\label{eq:Parseval}\end{equation}
Denote $\alpha_{j}=\left(y_{0},h_{j}\right)_{p},\quad\beta_{j}=\left(y_{1},h_{j}\right)_{p}.$
We  reduce the problem~(\ref{eq:_1})--(\ref{functional_u^2}) to
the following one:\begin{equation}
\int_{0}^{\infty}\sum s_{j}^{2}(t)dt\rightarrow\inf\label{functional_s^2}\end{equation}
\begin{equation}
\ddot{s_{j}}(t)+\lambda_{j}s_{j}(t)=C_{j}u(t),\quad j=1,2,\ldots\label{eq:s_j}\end{equation}
\begin{equation}
s_{j}(0)=\alpha_{j},\quad\dot{s}_{j}(0)=\beta_{j}\label{eq:nach_us_s}\end{equation}
\begin{equation}
-1\leq u(t)\leq1\label{upravlenie_q}\end{equation}
We shall assume everywhere below that \begin{equation}
C_{j}\neq0\quad\mbox{for all }j=1,2,\ldots\,\label{eq:Cj-not-0}\end{equation}
\smallskip{}
\textbf{Remark.} Assumption~(\ref{eq:Cj-not-0}) is very essential
for the problem~(\ref{functional_s^2})--(\ref{upravlenie_q}). Indeed,
let $C_{j_{0}}=0$ for some $j_{0}$. Then $j_{0}$-th equation in~(\ref{eq:s_j})
takes the form \[
\ddot{s}_{j_{0}}(t)+\lambda_{j_{0}}s_{j_{0}}(t)=0\]
 Hence, if $\left|\alpha_{j_{0}}\right|+\left|\beta_{j_{0}}\right|\neq0$
then the corresponding solution $s_{j_{0}}\left(t\right)$ does not
vanish as $t\rightarrow\infty$. Therefore the integral~(\ref{functional_s^2})
is equal to $+\infty$ and the optimization problem~(\ref{functional_s^2})--(\ref{upravlenie_q})
has not any sense.\smallskip{}

Assume that \begin{equation}
y_{0},\, y_{1}\in L_{2}\left(\left(0,l\right);p\right),\quad f\in L_{2}\left(0,l\right)\label{eq:L2}\end{equation}
 Following~\cite{bor_zel_manita_2008} we denote 

\[
\omega_{j}=\sqrt{\lambda_{j}},\quad\tau_{j}\left(t\right)=\dot{s}_{j}\left(t\right)/\omega_{j},\quad c_{j}=C_{j}/\omega_{j},\quad a_{j}=\alpha_{j},\quad b_{j}=\beta_{j}/\omega_{j}\]
Then the problem~(\ref{functional_s^2})--(\ref{upravlenie_q}) takes
the form\begin{equation}
\int_{0}^{\infty}\sum s_{j}^{2}(t)dt\rightarrow\inf\label{int_s}\end{equation}
\begin{equation}
\dot{s}_{j}=\omega_{j}\tau_{j},\quad\dot{\tau}_{j}=-\omega_{j}s_{j}+c_{j}u\label{eq:s_tau}\end{equation}
\begin{equation}
s_{j}(0)=a_{j},\,\,\tau_{j}(0)=b_{j},\,\, j=1,2,\ldots\label{na_us_s_tau}\end{equation}
\begin{equation}
-1\leq u(t)\leq1\label{eq:upravlenie_q}\end{equation}

Denote \[
s\left(t\right)=\left(s_{1}\left(t\right),s_{2}\left(t\right),\ldots\right),\quad\tau\left(t\right)=\left(\tau_{1}\left(t\right),\tau_{2}\left(t\right),\ldots\right)\]
\[
a=\left(a_{1},a_{2},\ldots\right),\quad b=\left(b_{1},b_{2},\ldots\right),\quad c=\left(c_{1},c_{2},\ldots\right)\]

\global\long\def\myargmax{\mathop{\mbox{arg\ensuremath{\,}max}}}

Consider the standart Hilbert space $l_{2}$: \[
l_{2}=\left\{ w=\left(w_{1},w_{2},\ldots\right):\,\,\, w_{n}\in\mathbf{R},\,\,\sum_{n=1}^{\infty}w_{n}^{2}<\infty\right\} \]
with inner product $\left(v,w\right)=\sum_{n=1}^{\infty}v_{n}w_{n}$.
Using assumption~(\ref{eq:L2}) we get that $a,\,\, b,\,\, c\in l_{2}$. 

The existence and uniqueness of a solution $\left(s\left(t\right),\tau\left(t\right)\right)$
to problem~(\ref{int_s})--(\ref{eq:upravlenie_q}) in the space~$l_{2}\times l_{2}$
were proved in~\cite{bor_zel_manita_2008} for any initial data from
an open neighborhood of the origin $\left(s=0,\tau=0\right)$. We
apply a formal generalization of the Pontryagin maximum principle
to the problem~(\ref{int_s})--(\ref{eq:upravlenie_q}). Denote by
$\psi_{i}=\left(\psi_{i1},\psi_{i2},\ldots\right)$ $(i=1,2)$ adjoint
variables. Define the Pontryagin function \begin{eqnarray*}
H\left(\psi_{1},\psi_{2},s,\tau,u\right) & = & \sum_{j=1}^{\infty}\left(\psi_{1j}\omega_{j}\tau_{j}-\psi_{2j}\omega_{j}s_{j}+\psi_{2j}c_{j}u-s_{j}^{2}/2\right)=\,\\
 & = & H_{0}\left(\psi_{1},\psi_{2},s,\tau\right)+uH_{1}\left(\psi_{1},\psi_{2},s,\tau\right)\end{eqnarray*}
where \[
H_{0}\left(\psi_{1},\psi_{2},s,\tau\right)=\sum_{j=1}^{\infty}\left(\psi_{1j}\omega_{j}\tau_{j}-\psi_{2j}\omega_{j}s_{j}-\frac{1}{2}s_{j}^{2}\right),\quad H_{1}\left(\psi_{1},\psi_{2},s,\tau\right)=\sum_{j=1}^{\infty}\psi_{2j}c_{j}\]
 For breavity we denote $z=\left(\psi_{1},\psi_{2},s,\tau\right)$.
In the space $l_{2}\times l_{2}\times l_{2}\times l_{2}$ let us consider
the Hamiltonian system \begin{equation}
\begin{array}{rclrcl}
\dot{\psi}_{1j} & = & \psi_{2j}\omega_{j}+s_{j},\quad & \dot{s}_{j} & = & \omega_{j}\tau_{j}\\
\dot{\psi}_{2j} & = & -\psi_{1j}\omega_{j}, & \dot{\tau}_{j} & = & -\omega_{j}s_{j}+c_{j}u^{*}(t)\end{array}\:\,\quad j=1,2,\ldots\label{eq:sopr_system}\end{equation}
where $u^{*}(t)$ satisfies the following maximum condition: \begin{equation}
u^{*}(t)=\myargmax_{u\in\left[-1,1\right]}H\left(z\left(t\right),u\right)=\myargmax_{u\in\left[-1,1\right]}\left(uH_{1}\left(z\left(t\right)\right)\right)\label{eq:optim_uprav}\end{equation}
Here we use notation:~ ${\displaystyle a^{*}=\myargmax_{a\in A}g(a)}$
iff ${\displaystyle g(a^{*})=\max_{a\in A}g(a)}$. 

It was proved~\cite{bor_zel_manita_2008} that the Pontryagin maximum
principle is the necessary and sufficient condition of optimality
for the problem~(\ref{int_s})--(\ref{eq:upravlenie_q}).

\newcommand{\abcd}{(\ref{eq:sopr_system})}

\global\long\def\HoSysSave{{\displaystyle \left.aH_{1}(t)\!\!\!\!\!\phantom{\int}\right|_{\abcd}}}
 \global\long\def\HoSys{{\displaystyle \left.\!\!\!\!\!\phantom{\int}\right|_{\abcd}}H_{1}(z)}
 \global\long\def\HoSysShort{\frac{d}{dt}{\displaystyle \left.\!\!\!\!\!\phantom{\int}\right|_{\abcd}\,}}
 \global\long\def\mysign{\mathop{\mbox{sign}}}

If $H_{1}\left(z\left(t\right)\right)\neq0$ along the trajectory
the optimal control is uniquely determined as a function of time from
the maximum condition (\ref{eq:optim_uprav}): \[
u^{*}\left(t\right)=\mysign\left(H_{1}\left(z\left(t\right)\right)\right)=\mysign\left(\sum_{j=1}^{\infty}\psi_{2j}(t)c_{j}\right)\]
 Suppose that there exists an interval $\left(t_{1},t_{2}\right)$
such that \[
H_{1}\left(z\left(t\right)\right)\equiv0,\quad\forall t\in\left(t_{1},t_{2}\right)\]
To find an optimal control $u\left(t\right)$ in this case we will
differentiate the identity $H_{1}\left(z\left(t\right)\right)\equiv0$
by virtue of the system~(\ref{eq:sopr_system}) until the control
$u$  with a non-zero coefficient occurs in the resulting expression
with a non-zero coefficient. \begin{eqnarray}
\frac{d}{dt}\HoSys & = & \HoSysShort\left(\sum_{j=1}^{\infty}\psi_{2j}c_{j}\right)=\left(-\sum_{j=1}^{\infty}c_{j}\psi_{1j}\omega_{j}\right)\nonumber \\
\frac{d^{2}}{dt^{2}}\HoSys & = & \HoSysShort\left(-\sum_{j=1}^{\infty}c_{j}\psi_{1j}\omega_{j}\right)=-\sum_{j=1}^{\infty}c_{j}\left(\psi_{2j}\omega_{j}^{2}+s_{j}\omega_{j}\right)\nonumber \\
\frac{d^{3}}{dt^{3}}\HoSys & = & -\HoSysShort\sum_{j=1}^{\infty}c_{j}\left(\psi_{2j}\omega_{j}^{2}+s_{j}\omega_{j}\right)=-\sum_{j=1}^{\infty}c_{j}\omega_{j}\left(-\omega_{j}^{2}\psi_{1j}+\tau_{j}\omega_{j}\right)\nonumber \\
\frac{d^{4}}{dt^{4}}\HoSys & = & \sum_{j=1}^{\infty}c_{j}\omega_{j}^{2}\left(\psi_{2j}\omega_{j}^{2}+2s_{j}\omega_{j}\right)-u\sum_{j=1}^{\infty}c_{j}^{2}\omega_{j}^{2}\label{eq:d4_H-1-1}\end{eqnarray}

Assume that all series in~(\ref{eq:d4_H-1-1}) are convergent in
$l^{2}$. Denote\[
H_{2}\left(z\right)=-\sum_{j=1}^{\infty}c_{j}\psi_{1j}\omega_{j},\quad H_{3}\left(z\right)=-\sum_{j=1}^{\infty}c_{j}\omega_{j}\left(\psi_{2j}\omega_{j}+s_{j}\right)\]
\[
H_{4}\left(z\right)=-\sum_{j=1}^{\infty}c_{j}\omega_{j}^{2}\left(-\psi_{1j}\omega_{j}+\tau_{j}\right)\]
 From~(\ref{eq:d4_H-1-1}) it follows that\[
H_{1}\left(z\left(t\right)\right)=H_{2}\left(z\left(t\right)\right)=H_{3}\left(z\left(t\right)\right)=H_{4}\left(z\left(t\right)\right)=0,\quad t\in\left(t_{1},t_{2}\right)\]
We say a solution of the (\ref{eq:sopr_system})--(\ref{eq:optim_uprav})
is singular if it belongs to the surface \begin{equation}
\Sigma=\left\{ \, z:\,\, H_{1}\left(z\right)=H_{2}\left(z\right)=H_{3}\left(z\right)=H_{4}\left(z\right)=0\,\right\} \label{eq:h_1-h_4-0}\end{equation}
 A singular control $u^{0}\left(t\right)$ is determited from the
equation $\frac{d^{4}}{dt^{4}}\HoSys=0$. Using~(\ref{eq:d4_H-1-1})
we obtain \begin{equation}
u^{0}(t)=\sum_{j=1}^{\infty}c_{j}\omega_{j}^{3}\left(\psi_{2j}\omega_{j}+2s_{j}\right)/\sum_{j=1}^{\infty}c_{j}^{2}\omega_{j}^{2}\label{eq:q^0}\end{equation}
Note that the origin $\left(\psi_{1}=0,\,\,\psi_{2}=0,\,\, s=0,\,\,\tau=0\right)$
is the singular trajectory and the corresponding singular control
$u^{0}\left(t\right)$ equals $0$. 

It was proved \cite{bor_zel_manita_2008} that in a certain neighborhood
of the origin the structure of the optimal solutions is the following
one: for the finite time the optimal nonsingular trajectory enters
the singular surface with infinite numbers of control switchings,
after that the optimal trajectory remains on the singular surface
and attains the origin for the infinite time. Namely, the following
theorem holds.

\begin{theorem} \cite{bor_zel_manita_2008}. \label{theorem_bor}
Let $c_{j}\neq0\quad\forall j$ and $\left(c_{1}\omega_{1}^{4},c_{2}\omega_{2}^{4},c_{3}\omega_{3}^{4},\ldots\right)\in l_{2}$.
Assume that there exist positive constants $\delta$ and $K$ such
that \[
\left|\omega_{j+1}\right|-\left|\omega_{j}\right|\geq\delta,\quad\left|\omega_{j}\right|\leq K\cdot j,\quad j=1,2,\ldots\]
 Then there exists an open neighborhood of the origin in the space
$\left(s,\tau\right)$ such that the following statements hold for
all initial data $\left(a,b\right)$ from this neighborhood. 

\begin{enumerate} \renewcommand{\labelenumi}{(\roman{enumi})}

\item  The problem~(\ref{int_s})--(\ref{eq:upravlenie_q}) has
a unique optimal solution.

\item  In the space $z=\left(\psi_{1},\psi_{2},s,\tau\right)$ there
exists the singular surface $\Sigma$ of codimension~4 given by the
equations \[
\sum_{j=1}^{\infty}\psi_{2j}c_{j}=0,\qquad\sum_{j=1}^{\infty}c_{j}\psi_{1j}\omega_{j}=0\]
\[
\sum_{j=1}^{\infty}c_{j}\omega_{j}\left(\psi_{2j}\omega_{j}+s_{j}\right)=0,\quad\quad\sum_{j=1}^{\infty}c_{j}\omega_{j}^{2}\left(-\psi_{1j}\omega_{j}+\tau_{j}\right)=0\]
 which is filled in by the singular extremals of the problem~(\ref{int_s})--(\ref{eq:upravlenie_q}).
The control on singular extremals are defined by~(\ref{eq:q^0}).

\item  For all initial data not belonging to the projection of the
singular surface $\Sigma$ on the space $\left(s,\tau\right)$, the
optimal trajectories arrive at $\Sigma$ in finite time with countable
many control switchings, i.e., the optimal trajectories are chattering
trajectories. 

\end{enumerate}

\end{theorem}

\section{Optimal solution for controlling vibrations }

Let $\left(s^{*}\left(t\right),u^{*}\left(t\right)\right)$ be an
optimal solution of~(\ref{int_s})--(\ref{eq:upravlenie_q}). Consider
\begin{equation}
y^{*}\left(t,x\right)=\sum_{j=1}^{\infty}s_{j}^{*}\left(t\right)h_{j}(x)\label{eq:rjad_u}\end{equation}
where $\left\{ h_{j}(x)\right\} _{j=1}^{\infty}$ are eigenfunctions
of the Sturm-Liouville problem~(\ref{eq:SLP})--(\ref{gran_usl_SLP}). 

Series~(\ref{eq:rjad_u}) formally satisfies equation~(\ref{eq:_1}),
boundary conditions~(\ref{eq:gr_1}) and initial conditions~(\ref{eq:na_us_1})--(\ref{na_us_2}).
We will show that this series gives a weak solution of problem~(\ref{eq:_1})--(\ref{na_us_2}).

Denote $Q_{T}=\left(0,l\right)\times\left(0,T\right)$, where $T>0$.
Consider the Sobolev space $H^{k}\left(Q_{T}\right)=W_{2}^{k}\left(Q_{T}\right),$
$k\geq0$. The space $H^{k}\left(Q_{T}\right)$ consists of all functions
$v\in L_{2}\left(Q_{T}\right)$ whose generalized derivatives up to
order $k$ exist and belong to $L_{2}\left(Q_{T}\right)$. The space
$H_{0}^{k}\left(Q_{T}\right)$ can be defined as a completion of $C_{0}^{\infty}\left(Q_{T}\right)$
with respect to the norm of the space $H^{k}\left(Q_{T}\right)$.

Let $g=uf\in L_{2}\left(Q_{T}\right),\quad y_{1}\in L_{2}\left(0,l\right)$.

\begin{definition} \tochka  We say a function $y\in H^{1}\left(Q_{T}\right)$
is a weak solution of the~(\ref{eq:_1})--(\ref{na_us_2}) if 

\begin{enumerate} \renewcommand{\labelenumi}{(\roman{enumi})}

\item ~ $\left.y\right|_{x=0}=\left.y\right|_{x=l}=0,\quad t>0$,
$\quad\quad\left.y\right|_{t=0}=y_{0}(x),\quad x\in[0,l]$;

\item~\[
\int_{Q_{T}}\left(ky_{x}v_{x}-py_{t}v_{t}\right)dxdt=\int_{Q_{T}}gv\, dxdt+\int_{0}^{l}y_{1}(x)v(0,x)\, dx\]
for each $v\in H^{1}\left(Q_{T}\right)$: $v\left|_{x=0}\right.=v\left|_{x=l}\right.=0,\, v\left|_{t=T}\right.=0$. 

\end{enumerate}

\end{definition}

\begin{definition} \tochka  We say $y\in H^{2}\left(Q_{T}\right)$
is a almost everywhere solution of the problem~(\ref{eq:_1})--(\ref{na_us_2})
provided $y$ satisfies equation~(\ref{eq:_1}) in $Q_{T}$ for almost
all $(t,x)\in Q_{T}$ and $y$ satisfies~(\ref{eq:gr_1})--(\ref{na_us_2}).

\end{definition}

The following theorem is the main result for the problem~(\ref{eq:_1})--(\ref{functional_u^2}). 

\begin{theorem} \tochka  \label{osn_theorem} Let $y_{0}\in H_{0}^{2}\left(0,l\right)$,
$y_{1}\in H_{0}^{1}\left(0,l\right)$, $k\left(x\right),p\left(x\right)$
are smooth enough (for example, $k,\, p\in C^{4}\left([0,l]\right)$),
$p\left(x\right)\geq p_{0}>0$, $k\left(x\right)\geq k_{0}>0$. Assume
that $f\in C^{4}[0,l]$, \begin{equation}
f(0)=f(l)=0,\quad f^{\left(i\right)}\left(0\right)=f^{\left(i\right)}\left(l\right)=0,\,\, i=1,2,3\label{usl_f(x)}\end{equation}
 and condition~(\ref{eq:Cj-not-0}) holds. Then there exist positive
constants $q_{1}$ and $q_{2}$ such that if \[
\left\Vert y_{0}\right\Vert _{L_{2}\left(\left(0,l\right);p\right)}<q_{1},\quad\left\Vert y_{1}\right\Vert _{L_{2}\left(\left(0,l\right);p\right)}<q_{2}\]
then \begin{enumerate} \renewcommand{\labelenumi}{(\roman{enumi})}

\item  the problem~(\ref{eq:_1})--(\ref{functional_u^2}) has a
unique optimal solution $y^{*}\left(t,x\right)$;

\item  $y^{*}\in H^{2}\left(Q_{T}\right)$ for all $T>0$; 

\item  an optimal solution $y^{*}\left(t,x\right)$ has an infinite
number of control switchings in a finite time interval. 

\end{enumerate}

\end{theorem}

\emph{Proof}. Here we use notations introduced in Section 2. Since
the functions $p,\, k,\, f$ satisfy conditions of Theorem~\ref{osn_theorem}
it follows \cite{Petrovsky} that $C_{j}\sim j^{-4}$ as $j\rightarrow\infty$,
where \[
C_{j}=\left(f,h_{j}\right)=\int_{0}^{l}f\left(x\right)h_{j}\left(x\right)dx\]
 Then we have \[
\sum_{j=1}^{\infty}\left(c_{j}\omega_{j}^{4}\right)^{2}=\sum_{j=1}^{\infty}\left(C_{j}\omega_{j}^{3}\right)^{2}=\sum_{j=1}^{\infty}C_{j}^{2}\lambda_{j}^{3}<\infty\quad\Longrightarrow\quad\left(c_{1}\omega_{1}^{4},c_{2}\omega_{2}^{4},c_{3}\omega_{3}^{4},\ldots\right)\in l_{2}\]
 The property~(\ref{asymp_lambda}) of the eigenvalues $\left\{ \lambda_{j}\right\} _{j=1}^{\infty}$
of the problem~(\ref{eq:SLP}) - (\ref{gran_usl_SLP}) imply that
there exist positive constants $\delta$ and $B$ such that \[
\left|\omega_{j+1}\right|-\left|\omega_{j}\right|\geq\delta,\quad\left|\omega_{j}\right|\leq Bj\]
 Now we may apply Theorem~\ref{theorem_bor} to the problem~(\ref{int_s})--(\ref{eq:upravlenie_q}).
We get that the optimal control $u^{*}\left(t\right)$ for the problem~(\ref{int_s})--(\ref{eq:upravlenie_q})
has an infinite number of switchings in the finite time interval.

Since the functions $f,\,\alpha,\,\beta$ satisfy the conditions of
Theorem~\ref{osn_theorem} it follows (see \cite{mihajlov,ladyzhenskaja})
that the function $y^{*}(t,x)$ defined by (\ref{eq:rjad_u}) is a
unique weak solution of the problem~(\ref{eq:_1})--(\ref{na_us_2})
and $y^{*}\in H^{2}\left(Q_{T}\right)$. Hence~\cite{mihajlov} $y^{*}(t,x)$
is the solution almost everywhere of the problem~(\ref{eq:_1})--(\ref{na_us_2}).
Thus the function $y^{*}(t,x)$ satisfies~(\ref{eq:_1}) for almost
all $(t,x)\in\left(0,l\right)\times\left(0,+\infty\right)$, boundary
and initial conditions~(\ref{eq:gr_1})--(\ref{na_us_2}). 

Since the function $s^{*}(t)=\left(s_{1}^{*}(t),s_{2}^{*}(t),\ldots\right)$
minimizes the functional~(\ref{functional_s^2}) and the identity~(\ref{eq:Parseval})
holds then the function $y^{*}\left(t,x\right)$ minimizes the func\-tio\-nal~(\ref{functional_u^2}).
Thus $y^{*}\left(t,x\right)$ is a solution of the problem~(\ref{eq:bound_control-1})--(\ref{functional_u^2}).

\section{Conclusion}

We considered  the optimal control problem of longitudinal vibrations
of a nonhomogeneous bar with clamped ends. We proved that the optimal
trajectories contain singular part and nonsingular one with accumulation
of control swithings.

~

\end{document}